# Nanostructured Europium Oxide thin films deposited by pulsed laser ablation of a metallic target in a He buffer atmosphere

H. Luna[1], D.F. Franceschini[2,a)], R. Prioli[3], R.B. Guimarães[2], C.M. Sanchez[2], G.P. Canal[4], M.D.L. Barbosa[5], R.M.O. Galvão[4]

[1] *Instituto de Física, Universidade Federal do Rio de Janeiro, Cx. Postal 68528, Rio de Janeiro, RJ 21941-972, Brazil*

[2] *Instituto de Física, Universidade Federal Fluminense, Niterói, RJ, 24210-346, Brazil.*

[3] *Departamento de Física, Pontifícia Universidade Católica do Rio de Janeiro, Rua Marques de São Vicente 225, 22453-970, Rio de janeiro, R.J., Brazil*

[4] *Centro Brasileiro de Pesquisas Físicas. Laboratório de Plasmas Aplicados, Rua Xavier Sigaud 150, 22290-180. Rio de Janeiro, R.J., Brazil*

[5] *Instituto de Física, Departamento de Física Nuclear, Universidade de São Paulo, Caixa Postal 66328, 05315-970, São Paulo, S.P., Brazil*

---

a) Email address: dante@if.uff.br (corresponding author)

Nanostrucured Europium oxide and hydroxide films were obtained by pulsed Nd:Yag (532 nm) laser ablation of an Europium metallic target, in the presence of a 1 mbar Helium buffer atmosphere. Both the produced film and the ambient plasma were characterized. The plasma was monitored by an electrostatic probe, for plume expansion in vacuum or in the presence of the buffer atmosphere. The time evolution of the ion saturation current was obtained for several probe to substrate distances. The results show the splitting of the plume into two velocity groups, being the lower velocity profile associated with metal cluster formation within the plume. The films were obtained in the presence of helium atmosphere, for several target to substrate distances. They were analyzed by Rutherford backscattering spectrometry (RBS), X-Ray Diffraction (XRD) and Atomic Force Microscopy, for as-deposited and $600^oC$ treated-in-air samples. The results show that the as-deposited samples are amorphous and have chemical composition compatible with Europium hydroxide. The thermally treated samples show X-Ray diffraction peaks of $Eu_2O_3$, with chemical composition showing excess oxygen. Film nanostructuring was shown to be strongly correlated to cluster formation, as shown by velocity splitting in probe current versus time plots.

## I. INTRODUCTION

Rare earth oxides have been widely used in luminescence devices, because of the electronic and optical characteristics originating from their 4f electrons[1]. The luminescence of $Eu^{3+}$ is particularly interesting because its major emission band is nearly centered at 611 nm wavelength (red), which is one of the primary colors. This leads to $Eu_2O_3$ being one of the most important oxide phosphors. To date, most of the work on Eu oxide phosphors was done on $Eu^{3+}$ doping of many host lattices, as a luminescence activator[2,3]. However, less attention has been dedicated to the study of pure Europium oxide as a phosphor. Recently, Yang *et al.* showed that nanostructuring Europium oxide could induce a strong red photoluminescence[4]. They produced Europium oxide nanotubes and nanorods, using a chemical route with carbon nanotubes as templates. Chemical routes were also employed to produce europium oxide nanoparticles, isolated[5] or immobilized on the surface of silica nanospheres[6], both showing also a remarkable red photoluminescence.

Despite its wide use in the production of metal oxide nanostructured materials and nanoparticles[7,8], laser ablation has been rarely used in the production of pure europium oxide nanostructured films. Concerning the use of laser ablation for the production of nanostructured europium oxide , most of the work was focused on the production of $Eu^{3+}$ doped metal oxides, mainly lanthanide oxides[8,9]. An exception to this is the production of nanoeuropium oxide ethanol sols, by laser ablation of an $Eu_2O_3$ target[10] submerged in a flowing liquid with chemical modifiers.

An interesting way to produce metal oxide nanoparticles or nanostrucutured films is to laser ablate a metallic target in the presence of a buffer inert gas atmosphere[11–14]. The interaction of the ablated plume with the background gas results in cluster growth, which can be revealed by the splitting of the plume fragments velocity profile[15], leading to the appearance of a lower velocity group, characteristic of the clustered atoms. The nanometric clusters condense onto the film growing surface, and oxidize when the deposition chamber is opened to the atmosphere. Such method was used to grow nanostructured tungsten oxide films, which were obtained with several morphology, from compact, nanostructured, or highly porous, depending on buffer gas, gas pressure and substrate-to-target distance[13,14]. Besides the advantage of the possibility of morphology control, this method is very suitable to europium oxide production, due the high reactivity of europium with oxygen, which may result in prompt oxidation of metal nanoparticles upon exposition to ambient atmosphere.

In this work, the results of a study on the production of nanostructured Europium oxide thin films by laser ablation of an Eu metallic target are reported in the presence of a He gas atmosphere. The variation of film structure, chemical composition and surface morphology are followed as a function of the target to substrate distance. In parallel, the velocity profile of the ablation plume ionic fragments is determined, in order to establish the correlation between plume expansion and film characteristics.

## II. EXPERIMENTAL PROCEDURE

The experimental apparatus used for film deposition is shown schematically in Fig.1. Briefly, a Q-switched Nd:YAG laser (Litron LPY 706-10 $^{TM}$), operating in the second harmonic ($\lambda$ = 532 nm), with 25 mJ pulse energy, is focused, at a 45 degree angle geometry, onto a pure metallic europium target surface placed inside a vacuum chamber ($5 \times 10^{-6}$ mbar without gas) generating a plasma that expands normally to the target surface ($z$ axis). The target holder is mounted on a rotating shaft controlled by a step motor, in order to expose a fresh surface after laser ablation. The laser and the step motor controller were synchronized in order keep a rate of 10 shots per degree. Care was taken to first clean any residual oxide layer from the europium by scanning the de-focused laser beam across the target surface.

The substrates, oxidized (600 nm thick $SiO_2$ layer) Si(100) chips, were placed symmetrically opposed to the target holder, and were mounted on a retractile holder in order to allow performing film depositions at different positions from the target surface. The ablation was performed at a constant helium pressure of 1 mbar, measured with a calibrated Pirani gauge. After each deposition the vacuum was broken and the metallic film exposed to air to react.

For the characterization of the plasma plume, a Langmuir probe is inserted at the position of the substrate holder, substituting it. The Langmuir probe was made of a 1.8 mm$^2$ copper disc and oriented to face the ablation spot, so that the plasma flux is perpendicular to the surface. With this arrangement, the expansion of the plasma along the z-axis was characterized[16,17].

The surface characterization was performed with an atomic force microscope (MultiMode, Veeco) operated in tapping mode and using a sharp silicon tip. A series of AFM images were obtained at each sample with scanning areas ranging from $1\mu^2$ to $100$ $\mu^2$. At least five different locations, $\sim 1$ mm apart from each odder, were analyzed and show the same morphology.

The chemical composition of the obtained films was monitored by Rutherford Backscattering Spectrometry (RBS). The RBS spectra were obtained with an impinging $2.2 MeV He^+$ beam, being the scattered beam detected at $170^o$ from the incident beam, by a surface barrier detector.

The charaterization of the crystal strucucture of the samples S30, S35 and S40 was done by X-ray diffraction at grazing incidence angles, in a Bruker D8 powder difractometer. The x-ray diffraction patterns were obtained using $CuK\alpha$ radiation, with $3^o$ fixed incidence angle, being the diffracted ray colimated by Soller slits and filtered by a LiF monochormator, before being detected by a scintilation counter.

## III. RESULTS: PLASMA ANALYSIS

In order to understand the expansion of the plasma plume, it is worthy to start from the analysis of free expansion, even considering that for deposition proposes; it does not configure the ideal condition. For the measurement of the ionic component of the plume, the probe was biased negatively -60 V, even though the ion saturation current sets in at voltages bellow -10V, in order to avoid plasma electrons and assure that the net current drawn by the probe is due only to the ion flux. Assuming that the ions are predominantly singly charged, due to the low laser fluency, the relation between the probe current and ion density is given by[18]:

$$I = A\, e\, n_i\, v, \qquad (1)$$

where $A$ is the probe area, $n_i$ is the ion density and $v$ is the ion velocity.

In figure 2 the probe ion current time of flight curve is shown for several positions along the expansion axis for free expansion (vacuum). Plotting the time of arrival of the peak of the saturation current as a function of position, along the expansion axis, yields a constant expansion velocity equal to $2.1 \times 10^6$ cm/sec. From equation 1, the variation of the maximum ion density can be calculated as a function of position, giving vales ranging from $9 \times 10^{13}$ ions/cm$^3$ at 10 mm to $2 \times 10^{12}$ ions/cm$^3$ at 40 mm.

A second possible deposition configuration is the expansion of the plasma plume into a high pressure gas atmosphere[21]. In this work we chose helium as background gas because of its high ionization energy (25 eV), hence avoiding ionization. The working pressure was kept constant at 1 mbar. The idea is to create a deposition condition in which a phenomena known as plume splitting may occur.

Plume splitting of a laser produced plasma has been studied using Langmuir probes[15,19,20,22] and also by a fast imaging technique[23–25]. Briefly, ablation into ambient gases results in shock waves and expansion fronts propagating through the background gases. Once occurring the splitting of the plasma plume, there is an energetic component that propagates with a free expansion velocity and a second one that is slowed down, depending on the pressure of the background gas. Therefore, the ions located at the front of the plasma acquire the largest energy during hydrodynamic acceleration and the interaction time for recombination is very much reduced. On the other hand, the ions located in the inner plume layers are less energetic. They remain much longer in the denser state, which may be subjected to strong recombination. Recently, Kushwaha and Thareja performed an elegant work where the splitting of a carbon plasma plume expanding in a nitrogen gas atmosphere was recored using a time resolved spectroscopy imaging technique[26]. They found that, at early stages of the plume expansion, the outer plume (energetic) yields a spectroscopic emission dominated by CI and CII species, while for the inner plume the spectrum was almost entirely dominated by $C_2$ and CN band line emission. Therefore the outer plume, under proper conditions, might be the right ambient for clustering followed by subsequent nanoparticle formation and film growth[27].

With the purpose of better understanding the role of the plume splitting in the formation of nanostructures, we have studied the spatial evolution of the europium plume at a 1 mbar helium atmosphere. In figure 3 the probe current time of flight curve is shown for 4 distances, namely $z = 15, 30, 35$, and 40 mm. promptly see that at 15 mm the plume has not split yet. From 30 mm, however, a small inner plume starts to appear. At 40 mm a total split between a outer and an inner plume is evident, with the magnitude of the ion current from inner plume taking over the outer plume.

A position-time plot of the maximum ion current is shown in figure 4 for the outer (full square) and for the inner plume (full circle). The plume expansion can be described by a classic drag model[20,24] in which the plasma plume is regarded as an ensemble that experiences a viscous force proportional to its velocity through the background gas. Therefore the position and velocity are given by $Z = Z_f(1 - exp(-\beta t))$ and $V = V_0 \, exp(-\beta t)$, respectively. For the position there are two fitting parameters $Z_f$ (stopping distance of the plume) and $\beta$ (slowing coefficient). The drag model is also shown for the outer plume in figure 4 (continuous line) with $Z_f = 43.8$ mm and $\beta = 0.47$ $\mu sec^{-1}$. Although the drag model fits the outer plume reasonably well, it predicts that the plume will eventually stop at a distance $Z_f$ of 43.8 mm, which is obviously in disagreement with the experimental observation.

It is interesting to note that for the inner plume the drag model fails to fit the expansion. The numerical derivative of the position-time plot of figure 4 yields velocities decreasing from $1 \times 10^6$ cm/sec to $0.4 \times 10^6$ cm/sec for the outer plume, and $0.4 \times 10^6$ cm/sec to $0.1 \times 10^6$ cm/sec for the inner plume.

The Langmuir probe was then replaced and the substrate holder with the substrate placed at distances, $z = 15, 30, 35,$ and 40 mm. The film morphology will be discussed in the following section.

## IV. RESULTS: FILM ANALYSIS

### A. Atomic Force Microscopy

Four different film preparations were studied in this work. They were obtained with 15, 30, 35, and 40 mm target-to-substrate distances, and will be referred as samples S15, S30, S35 and S40. The S15 and S35 samples were analyzed as-deposited. The S30 and S40 samples were thermally treated at 600 $^oC$ for five hours in air before being analyzed. The AFM images of the samples are shown in Figs. 5-8.

Sample S15 presented in Fig. 5 shows a very flat surface, partially covered by particulates with a wide diameter distribution, being some of them rather large, with more than one micrometer. Therefore, we attribute the presence of these particulates at the surface to the splash of target material by the laser. Particulates generated by pulsed laser ablation can be classified into three categories. A specific type of particulate is distinguished by whether the original matter, right after being ejected from the target, is in solid, liquid or vapor phase. Briefly, the size of the paticulate when originated from vapor phase should be in the nanometer range, while the two other cases should be in the sub-micron to micron range.

Laser wavelength also plays an important role in the formation of particulates because of the effectiveness of the absorption of the laser power into the target[30]. It is well know that for the wavelengths of 1064 nm and 532nm the particulate formation is mainly dominated by massive splashes of the target material, for irradiances above 0.1 GW/cm$^{-2}$[31].

The AFM images of the other samples presented in Figs.6-8 also show splash particulates originated from the splash of the target but, with lower surface densities and sizes. Despite the particulates, the topography of samples S30-S40 show a degree of nanostructuring, which depends on substrate-to-target distance and the processing history. In order to take into account the processing history we must first compare the as-deposited samples and the thermally processed samples separately.

The as-deposited samples show different surface topographies. While sample S15 (Fig. 5 exhibits a flat and compact surface, sample S35 (Fig. 7) is found to have a clearly nanostructured surface. Similar comparison can be made between the 600 $^oC$ treated samples, S30 and S40, shown in figures 6 and 8. Despite being also nanostrucutured, sample S30 is more compact then S40, which shows structures with larger dimensions and a wider size distribution.

Complementing these data, Fig. 9 shows higher resolution profiles taken from sample S30. The surface of sample S30, shown in Fig. 9 (a) is composed by particles with diameter ranging from 26 to 160 nm. In Fig. 9(b) the profile obtained from the bottom of a crater is shown to be similar to the film surface. The crater was formed by the removal of a particle splashed from the target. Thus, we can imagine that the film as a whole is composed by particles similar to that present on the surface.

The AFM results show that the surface texture, or the particle size distribution, strongly depends on the target-to-substrate distance. It is expected that the velocity distribution of the ions in the plasma plume must be correlated with clustering effects, in particular to the inner plume composed by the lower velocity group. As shown in figure 3, as the distance relative to target increases, the peak velocity of the inner plume (slower group) decreases while its integrated peak intensity increases relative to the outer plume (faster group).

One can observe that the S15 sample, which does not show surface nanostructuring effects at all, is obtained with a target-to-substrate distance bellow the one corresponding to plume splitting shown in 3. Taking separately the as-deposited and heat treated samples, one can observe that the particle size, or the nanostrucuture dimension, increases with target-to-substrate distance. Even considering the different processing conditions of the samples, our results suggest that the particle size increases continuously as the inner plume (slower group) takes over the outer plume (faster group). The SEM images of $WO_x$ films reported by Bailini et. al.[14], obtained with the same deposition method used in this work, suggest that there is a continuous decreasing of compact mater as the ratio of the target-to-substrate distance to the time integrated visible plume length is increased. This ratio, is almost equivalent to the target-to-substrate distance variation in present work, since we have not changed pressure or gas atmosphere.

## B. Rutherford Backscattering Spectrometry

The four samples studied in this work were submitted to RBS analysis. The RBS spectra are shown in Fig. 10, together with the RUMP simulation[28] used for chemical composition determination. The Europium and Oxygen surface backscattering edges are indicated in the spectrum of sample S15. For the outermost layer (the film itself), spectra show only the presence of oxygen and europium. The film chemical composition was shown to be uniform upon depth variation, since there is a good agreement between spectra and simulation, which was based on uniform composition of each layer. The spectra show also Si and O in innermost layers ($SiO_2$ layer and $Si$ substrate). To discuss the chemical composition results, we will first consider separately the as-deposited and heat-treated samples. The results of chemical composition of all deposited films are displayed in I, together with the film thickness.

Beginning with the as-deposited samples, we see that the oxygen content is 76.7 % for sample S15, which is equivalent to O:Eu ratio near 3:1. This ratio is higher than any Europium Oxide stoichiometry, and seems to correspond to Europium Hydroxide - $Eu(OH)_3$ or $Eu(OH)_2.2(H_2O)$. Since RBS cannot detect hydrogen atoms, we cannot say with confidence that there is hydrogen present in the film. On the other hand, it is well known that metallic filamentary Europium reacts with moist air, being completely converted in europium hydroxide within hours[29]. We expect that an metallic europium thin film would suffer the same transformation to europium hydroxide. The S35 sample, also as-deposited, shows an even higher oxygen content of 80.8 %. This high value may be a consequence of the presence of water molecules trapped by the porosity induced by the nanometric dimensions of the deposited clusters.

The heat treated samples, S30 and S40, show a lower oxygen content than the untreated sample S35. Sample S30 shows an oxygen content equal to 78.3 %, which is still very near to the 80.3 % of the untreated S35 sample, which suggests an incomplete dehydration of S30 sample, even after being heated at 600 ºC. Sample S40 shown a 68.7 %, which seem to approach to the 60 % value of the stoichiometric $Eu_2O_3$, the phase expected to be obtained from hydroxide heat treatment in air. At the moment, we have no explanation for this high oxygen content, which deserves further investigation.

## C. X-Ray Diffraction

The X-ray diffraction patterns obtained from samples S30, S35 and S40 are shown in Fig. 11. Sample S35, which was analyzed as deposited, shows an amorphous pattern, similar to that obtained by Haopeng et al. after drying Europium hydroxide coated carbon nanotubes[4]. Difractograms from samples S30 and S40, which were heat treated, show clearly developed $Eu_2O_3$ diffraction lines, which are indexed in the figure. The two difractograms are nearly identical and have broad lines (about 0.5 degrees), corresponding to small crystals. Again, the diffraction patterns are very similar to that obtained by treating at 600 ºC in air, carbon nanotubes coated with europium hydroxide in ref.[4].

## V. CONCLUSION

Nanostrucured Europium oxide or hydroxide films were obtained by pulsed Nd:Yag (532 nm) laser ablation of an Europium metallic target, in the presence of a 1 mbar Helium buffer atmosphere. Film and plasma were both characterized. The plasma was monitored by an electrostatic probe, for plume expansion in vacuum or in the presence of the buffer atmosphere. Current versus time plots were obtained, for several probe to substrate distances. The results show the splitting of the plume into two velocity groups, being the lower velocity profile associated with metal cluster formation within the plume. The films were obtained in a helium atmosphere for several target to substrate distances and were analyzed by Rutherford backscattering spectrometry (RBS), X-Ray Diffraction (XRD) and Atomic Force Microscopy, for as-deposited and 600 $ºC$ treated-in-air samples. The results show that the as-deposited samples are amorphous and have chemical composition compatible with Europium hydroxide. The thermally treated samples show X-Ray diffraction peaks of $Eu_2O_3$, with chemical composition showing excess oxygen. Film nanostructuring was shown to be strongly correlated to cluster formation, as shown by velocity splitting in probe current versus time plots.

**Acknowledgements** This work was partially supported by CNPq - Conselho Nacional de Desenvolvimento Científico e Tecnlógico. The authors are thankful to Dr. Ricardo Castell for useful discussions on the laser ablation technique.

TABLE CAPTIONS

Table 1-RBS results

FIGURE CAPTIONS

Figure 1 - Sketch of the experimental set-up for pulsed laser deposition.

Figure 2 - Probe current time of flight curve for the free expansion.

Figure 3 - Probe current time of flight curve for the expansion at 1 mbar of helium atmosphere.

Figure 4 - Position-time plot of the maximum ion current for the expansion at 1mbar of helium atmosphere. Open square - plume before splitting; Full square - outer plume (faster group velocity); Full circle - inner plume (slower group velocity); and continuous line - Drag model fitting for the outer plume.

Figure 5 - Surface topography of sample S15 as determined by Atomic force Microscopy.

Figure 6 - Surface topography of sample S30 (thermally treated) as determined by Atomic force Microscopy.

Figure 7 - Surface topography of sample S35 as determined by Atomic force Microscopy.

Figure 8 - Surface topography of sample S40 (thermally treated) as determined by Atomic force Microscopy.

Figure 9 - Enlarged surface topography of sample S30 as determined by Atomic force Microscopy.

Figure 10 - $^4He$ RBS spectra of all samples.

Figure 11 - X-Ray diffraction paterns from samples S30, S35 and S40.

TABLE I.

| Sample/ Processing | Thickness $(10^{15})$ atoms/$cm^3$ | Eu content at. % | O content at. % |
|---|---|---|---|
| S15 as deposited | $6670 \pm 50$ | $23.3 \pm 0.7$ | $76.7 \pm 2.3$ |
| S30 $600^oC$ treated | $1070 \pm 50$ | $21.7 \pm 0.7$ | $78.3 \pm 2.3$ |
| S35 as deposited | $2150 \pm 50$ | $19.2 \pm 0.6$ | $80.8 \pm 2.4$ |
| S40 $600^oC$ treated | $1100 \pm 50$ | $31.3 \pm 0.9$ | $68.7 \pm 2.1$ |

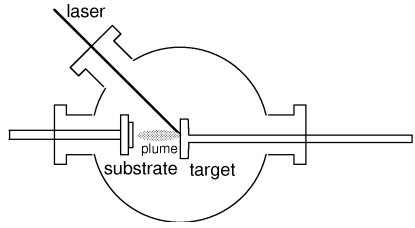

FIG. 1.

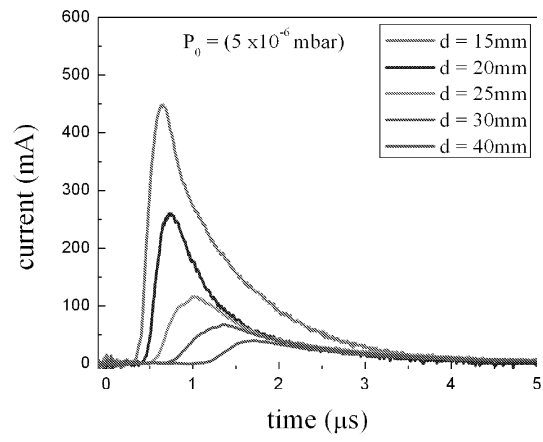

FIG. 2.

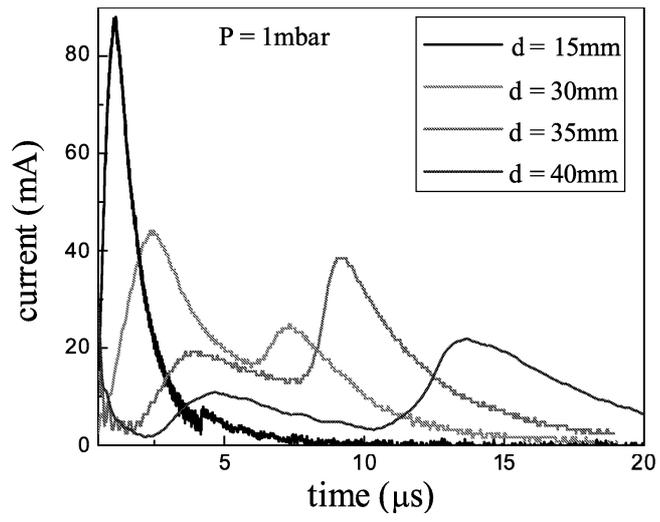

FIG. 3.

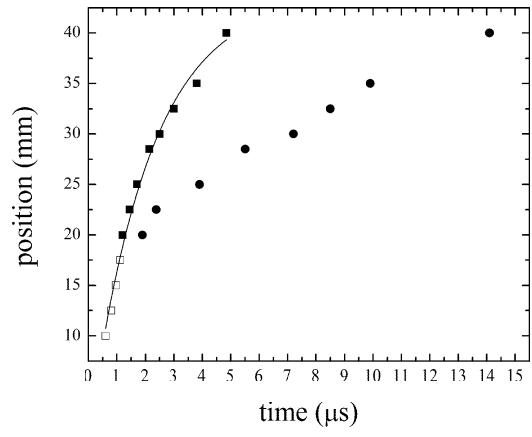

FIG. 4.

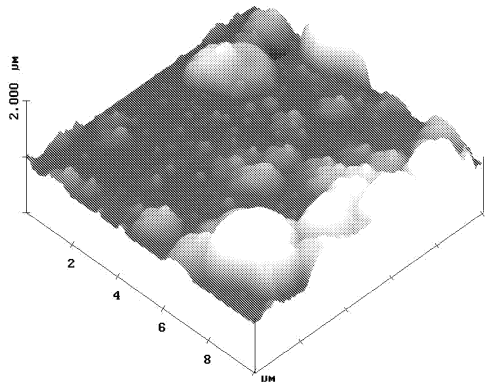

FIG. 5.

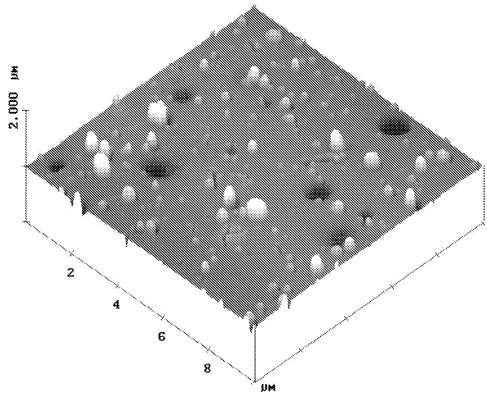

FIG. 6.

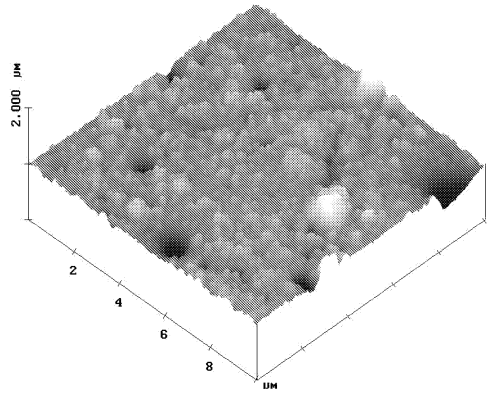

FIG. 7.

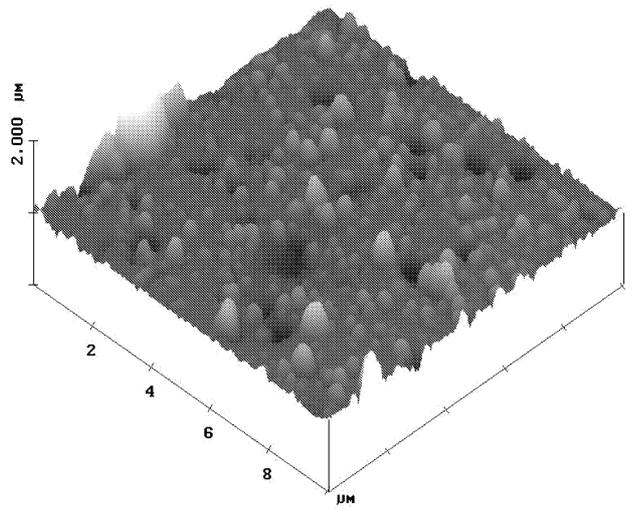

FIG. 8.

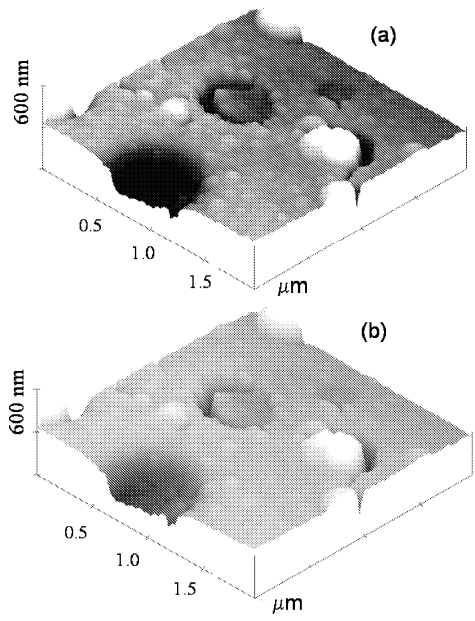

FIG. 9.

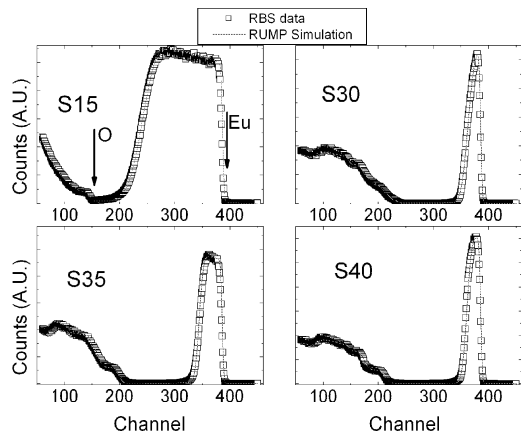

FIG. 10.

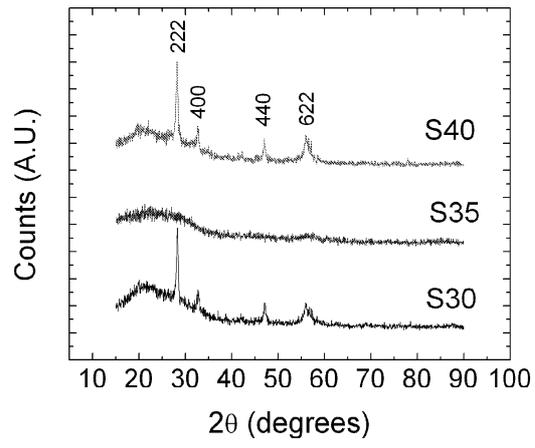

FIG. 11.